\documentclass[lettersize,journal]{IEEEtran}
\usepackage{amsmath,amsfonts}
\usepackage{algorithmic}
\usepackage{algorithm}
\usepackage{array}
\usepackage[caption=false,font=small,labelfont=rm,textfont=rm]{subfig}
\usepackage{textcomp}
\usepackage{stfloats}
\usepackage{url}
\usepackage{verbatim}
\usepackage{color}
\usepackage{graphicx}
\usepackage{booktabs}
\usepackage{cite}
\usepackage{indentfirst}
\begin{document}

\title{Realizing Immersive Communications in Human Digital Twin by Edge Computing Empowered Tactile Internet: Visions and Case Study}
\author{Hao Xiang, Changyan Yi,~\IEEEmembership{Member,~IEEE,} Kun Wu, Jiayuan Chen, Jun Cai,~\IEEEmembership{Senior Member,~IEEE,}
\\Dusit Niyato,~\IEEEmembership{Fellow,~IEEE} and Xuemin (Sherman) Shen,~\IEEEmembership{Fellow,~IEEE}
 \thanks{H. Xiang, C. Yi, K. Wu and J. Chen are with the College of Computer Science and Technology, Nanjing University of Aeronautics and Astronautics, China. (Email: xianghao@nuaa.edu.cn, changyan.yi@nuaa.edu.cn, wukun5@nuaa.edu.cn, jiayuan.chen@nuaa.edu.cn).

J. Cai is with the Department of Electrical and Computer Engineering, Concordia University, Montreal, Canada. (Email: jun.cai@concordia.ca).

D. Niyato is with the School of Computer Science and Engineering, Nanyang Technological University, Singapore (Email: dniyato@ntu.edu.sg).

X. Shen is with the Department of Electrical and Computer Engineering, University of Waterloo, Canada. (Email: sshen@uwaterloo.ca).}
}



\maketitle

\begin{abstract}
Human digital twin (HDT) is expected to revolutionize the future human lifestyle and prompts the development of advanced human-centric applications (e.g., Metaverse) by bridging physical and virtual spaces. However, the fulfillment of HDT poses stringent demands on the pervasive connectivity, real-time feedback, multi-modal data transmission and ultra-high reliability, which urge the need of enabling immersive communications. In this article, we shed light on the design of an immersive communication framework for HDT by edge computing empowered tactile Internet (namely IC-HDT-ECoTI). Aiming at offering strong interactions and extremely immersive quality of experience, we introduce the system architecture of IC-HDT-ECoTI, and analyze its major design requirements and challenges. Moreover, we present core guidelines and detailed steps for system implementations. In addition, we conduct an experimental study based on our recently built testbed, which shows a particular use case of IC-HDT-ECoTI in physical therapy, and the obtained results indicate that the proposed framework can significantly improve the effectiveness of the system. Finally, we conclude this article with a brief discussion of open issues and future directions.
\end{abstract}
\section{Introduction}\label{Introduction}


\IEEEPARstart {H}UMAN digital twin (HDT)  is envisioned to perform as a hyper-realistic and hyper-intelligent testbed for far-reaching fields of human-centric services (such as Metaverse) \cite{ref2}. HDT for each individual consists of a pair of physical twin (PT) and virtual twin (VT). Specifically, the VT in the digital space can replicate its corresponding human body in the physical space, i.e., PT, and can also reflect the PT's status both psychologically and physiologically in real time. For example, HDT may be implemented to plan and assist complex and high-risky surgeries, where doctors can remotely observe patients' physiological indicators, operate medical equipments, conduct accurate treatments and perform various tests via physical-virtual synchronization. HDT obviously requires strong interactions and extremely immersive quality of experience (QoE) among PTs and VTs, urging the need of enabling immersive communications \cite{refadd1}.

Edge computing empowered tactile Internet (ECoTI), by integrating various sensors with communication and computational capabilities at network edges, can transmit human skills through networks and provide multisensory haptic feedbacks, allowing users to interact with objects more intuitively and more fast-responsively. With these features, ECoTI may be a promising solution to realize immersive communications for HDT. Hereafter, for convenience, we call this new framework as IC-HDT-ECoTI.
\begin{enumerate}
  \item [a)] \emph{Realizing immersive communications in HDT is challenged due to the frequent service interruptions caused by unpredicted human activities and limited resources.} IC-HDT-ECoTI can potentially address this issue by providing pervasive and agile connections. On one hand, it can support large-scale data collections, real-time data processing and analysis, enabling data-driven HDT applications. On the other hand, it can employ distributed approaches to optimize resource allocations, improving the availability, scalability and fault tolerance.
  \item [b)] \emph{Realizing immersive communications in HDT is difficult when facing heavy and fluctuating traffics, inducing delays and inconsistencies in feedbacks across different data modalities (e.g., video, audio and tactile).} IC-HDT-ECoTI can potentially address this issue by improving network efficiencies and qualities. Particularly, data processing and analysis can be transferred from the cloud to edge servers closer to data sources and terminals. This reduces transmission distances, ensuring the effective management of HDT while maintaining information consistency, integrity and security.
  \item [c)] \emph{Realizing immersive communications in HDT requires high-fidelity virtual-real interactions while PT-VTs are hard to be seamlessly exchanged and synchronized.} IC-HDT-ECoTI can potentially address this issue by building systems over multiple dimensions. In the modeling dimension, it can manage, analyze, mine and integrate multi-source data collected from physical entities, enabling real-time updates and interactions. In the service dimension, it can achieve high-precise simulation, providing desired functions and services, e.g., human activity detection and healthcare analysis, according to specific HDT user requirements.
\end{enumerate}

\par In summary, to be capable of offering strong interactions and extremely immersive QoE in facilitating advanced human-centric applications, IC-HDT-ECoTI is not only attractive but also worthy to be carefully studied. The main contributions of this article are to provide a bold, forward-looking vision on designing IC-HDT-ECoTI, and are summarized as follows.

\begin{itemize}
\item We propose a novel communication framework, i.e., IC-HDT-ECoTI, aiming to provide HDT with strong interactions and extremely immersive QoE services. To the best of our knowledge, this is the first framework that manages holistic HDT functions empowered by IC-HDT-ECoTI, including data collection, processing and transfer.
\item We rigorously analyze key design requirements and challenges of implementing IC-HDT-ECoTI, and present the key steps and core guidelines.
\item We show a particular case study of IC-HDT-ECoTI in physical therapy, demonstrating the effectiveness of the proposed framework. This use case may also be considered as a reference for other HDT applications that can improve human quality of life.
\item We outline and discuss major open issues to inspire future research directions.
\end{itemize}

\section{Framework of IC-HDT-ECoTI}\label{Framework}

\subsection{System Architecture}

\begin{figure*}[t]
	\centering
	\includegraphics[width=0.87\linewidth]{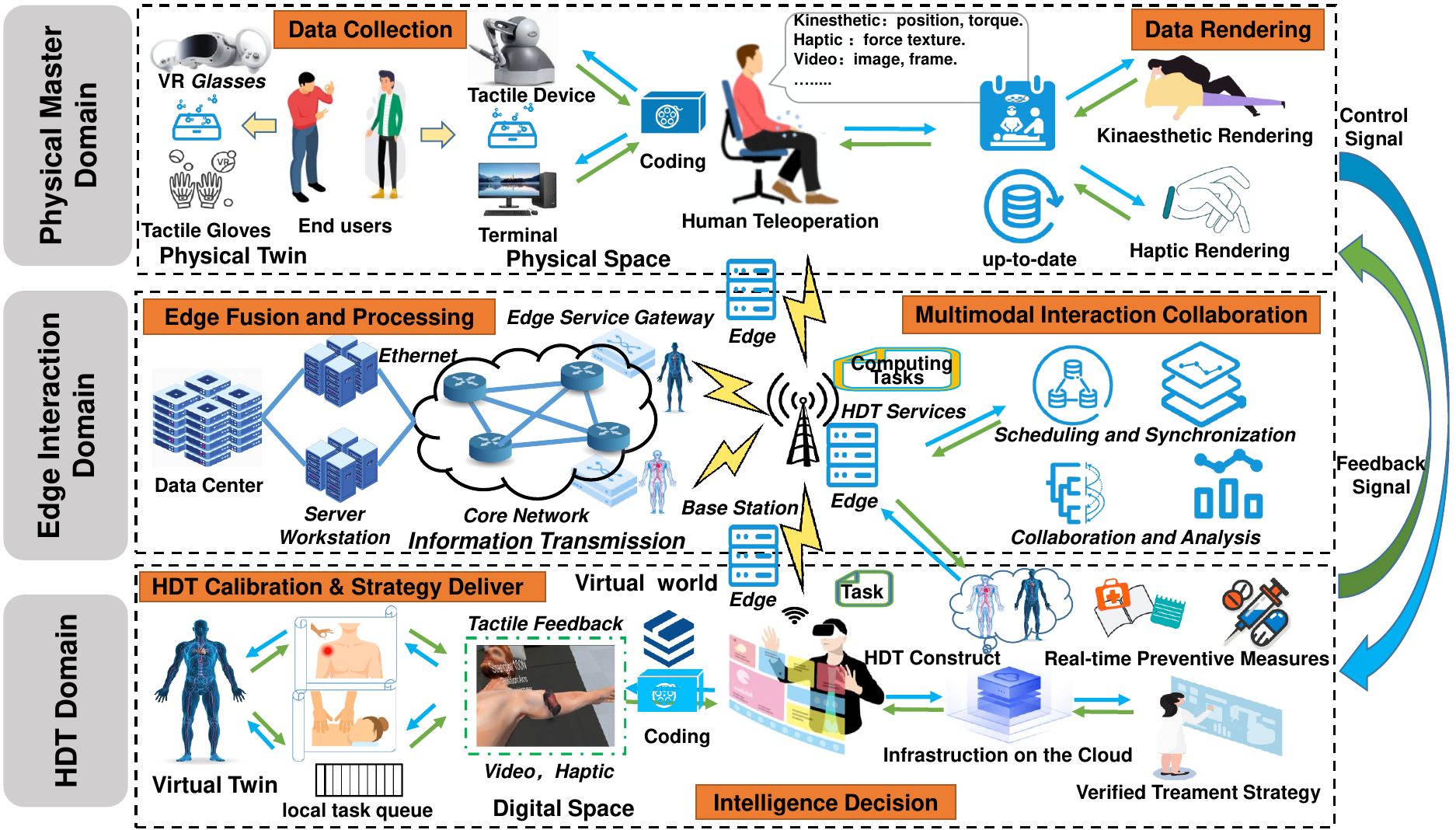}
	\caption{The system architecture of IC-HDT-ECoTI, including the physical master domain, edge interaction domain and HDT domain.}
	\label{fig1}
\end{figure*}

The system architecture of IC-HDT-ECoTI consists of three domains, i.e., physical master domain, edge interaction domain and HDT domain, as illustrated in Fig. \ref{fig1}. PTs in the physical master domain (i.e., physical space) can initiate human skills, such as massage or acupuncture through human teleoperators and tactile terminals, to control VTs in the HDT domain (i.e., digital space), and obtain ultra-realistic feedback from the HDT domain. As the bridge, the edge interaction domain helps deliver commands and feedback signals between the physical master and HDT domains in real-time.

\subsubsection{Physical Master Domain}
\par Broadly speaking, PTs in the physical master domain can be equipped by various haptic sensors, tactile terminals and display devices to interact with VTs in the HDT domain to complete specific applications. For example, a doctor equipped with tactile gloves can control his/her VT (i.e., the virtual hand) to massage the patient's VT within the HDT domain. The tactile feedback will be encoded and transmitted back to the doctor through the edge interaction domain. In addition, tactile, video and kinesthetic information may be imperative in this scenario, and thus motion actuators and head-mounted VR glasses may also be required to build a two-way round-trip operation framework.

\subsubsection{Edge Interaction Domain}
\par The edge interaction domain supports bidirectional communication links between the physical master domain and HDT domain. Besides the capability of information exchange, edge computing can promote the data fusion and reduction, supporting complex HDT tasks. Additionally, synchronizing multi-modal feedback by analyzing the data at the network edge can further bring an immersive multi-sensory experience to PTs.

\subsubsection{HDT Domain}
\par A PT is digitally modeled as a high-fidelity VT located in the HDT domain, which can be regarded as the digital executor of the PT, meanwhile reflecting the PT's status in real time. After receiving
commands from the PT, its VT decodes the information and executes actions indicated by the decoded information. Based on this, the VT then generates the multimodal feedback information and transmit it back to the PT through the edge interaction domain.

\subsection{Key Design Requirements and Challenges}\label{KD}
The key design requirements and major challenges of implementing IC-HDT-ECoTI are discussed as follows.

\subsubsection{Strong Interactions between Physical and Virtual Twins}
The interactions between a pair of PT and VT typically involve information exchange in a high frequency. For instance, for the smoothness and fidelity of haptic perception when a PT switches actions rapidly, haptic information commonly needs to be updated at a rate over 1000 times per second. These strong interactions between PTs and VTs need the support of pervasive connectivity under ubiquitous mobility, real-time communication and computation with feedbacks, and privacy protection and data security for ethics and morality.

\par \textbf {Pervasive Connectivity under Ubiquitous Mobility}: The interactions between PTs and VTs through IC-HDT-ECoTI inherently demand seamless connectivity. HDT services may be interrupted when a PT moves to distant locations, while its VT is still hosted at the original server. Hence, the pervasive connectivity is required for guaranteeing stable and continuous services \cite{ref5}. Nevertheless, this requirement brings several challenges. First, when the PT moves to a place with unpredictable network conditions, such as high congestions and packet loss rates, the connectivity between the PT and VT pair is hard to be established. Second, if the PT moves from an area with a low user density to other high-density ones, the network workload will become more unbalanced, and the topology will become more complex to manage. Third, the mobility of PT is significantly uncertain, depending on their subjective consciousness. All these indicate that the pervasive connectivity of PT-VT pair under ubiquitous mobility is challenging.

\par \textbf {Real-time Communication and Computation with Feedbacks}: Intuitively, strong interactions between PTs and VTs require the support of real-time communications and computations. First, feedbacks from VTs typically involve massive complex and multimodal information (e.g., audio, video and tactile signals) that requires large bandwidth to be transmitted, and thus the network resource needs to be optimized. However, most of existing work devoted to the design of uplink task offloading while ignoring the downlink feedbacks \cite{ref6}, which results in the low quality of response. Second, if the distance between physical master domain and HDT domain is more than 150 kilometers, it is generally known that real-time communication requirements can hardly be achieved \cite{ref7}. This limitation hinders the application of IC-HDT-ECoTI, such as for ultra-long-distance surgery. Moreover, edge computing nodes usually have limited computing capacity, storage space, battery life and other resource constraints, and may be seriously affected by network congestions, equipment failures and attacks. These motivate the enhancement of availability, reliability, scalability and fault tolerance of edge services.

\par \textbf {Privacy Protection and Data Security with Ethics and Morality}: Massive data will be transmitted and exchanged between PTs and VTs in IC-HDT-ECoTI during the strong interactions. Privacy protection and data security are undoubtedly required for preventing data from malicious leakages or attacks. Different from the other general applications, the data generated in IC-HDT-ECoTI are human-related and highly sensitive. Any leakage of these data may result in serious ethical and moral concerns. For example, in the treatment of patients under such a framework, a large amount of  healthcare data may be circulated. Doctors taking actions and interacting with patients' VTs  is vulnerable to malicious attacks on data transmission, processing and feedbacks, causing potential data manipulations and eventually the mismatch among PTs and VTs. These attacks may further lead to misjudgement, wrong decisions and failure of treatments, which are intolerable in any case \cite{ref8}. Moreover, the traditional data backup, recovery, auditing schemes may not be sufficient for this application, because the operation of IC-HDT-ECoTI has to maintain relatively stable with few jitters and service interruptions.

\subsubsection{Extremely Immersive Quality of Experience}
\par The extremely immersive QoE of IC-HDT-ECoTI can be achieved by providing PTs with comprehensive and ultra-realistic virtual senses and actions generated by VTs. This requires the support of high fidelity virtual modeling, multimodal data analysis, and the integration of subjective and objective evaluations.

\par \textbf {High Fidelity Virtual Modeling}: High-fidelity virtual scenes of objects, tasks and actions interconnecting the physical master domain and HDT domain have to be established for IC-HDT-ECoTI between PTs and VTs. This helps ensure interactive effects in the design of prototypes, while guaranteeing  immersive QoE for users. However, the realization of such high-fidelity virtual modeling brings several challenges. First, most of the existing tactile devices have not been fully developed to capture ultra-fine-grained information, and may sometimes capture inaccurate or wrong information, making the immersive QoE being significantly degraded \cite{ref9}. Second, the current virtual scene modeling with offline preprocessing cannot be easily migrated from one server to another, meaning that it is not applicable for high mobility services with quick motion changes or scene switches. Last but not least, there is an inherent tradeoff between the sampling frequencies of tactile devices and the communication traffics. To be more specific, the excessive sampling frequency (e.g., 1 kHZ in the remote surgery \cite{ref15}) may lead to considerably high traffics occupying large communication bandwidths, and thus increasing the service latency due to potential network congestions. In contrast, the over-low sampling frequency may inevitably result in ``unrealistic feeling''.


\par \textbf {Multimodal Data Analysis}: Multimodal data is the fuel of IC-HDT-ECoTI, which normally contains various types of data from multiple sources. However, the transmission and processing of such data bring several open problems. First, multimodal encoding schemes must be defined to support different modalities, without compromising the end-to-end service latency. Currently, although visual and auditory information encoding has been studied extensively, haptic encoding is still very challenging \cite{ref10}. Second, the processing of multimodal data needs to be synchronous. Otherwise, cybersickness may happen. Cybersickness commonly refers to an issue that multiple sensing data (e.g., audio, video and tactile sensation) are incorporated in an interaction but their data processing are notably asynchronous (e.g., the time-lag between tactile and audio movement exceeds 1 ms) \cite{ref11}.

\par \textbf {Integration of Subjective and Objective Evaluations}: Unlike the conventional  systems, IC-HDT-ECoTI requires both objective and subjective metrics to evaluate the performance, including not only the latency, reliability of communication and computation, but also the true experience in immersive interactions. However, it is difficult, if not impossible, to design an integrated evaluation method that can jointly capture both subjective and objective aspects, as it may involve interdisciplinary knowledge and complex models. Furthermore, compared to the objective evaluation, it is worth noting that users' feedbacks (e.g., error corrections) in subjective evaluation require much longer response time, making the real-time integration and analysis much more challenging \cite{ref12}.

\section{Key Steps and Core Guidelines}\label{Kep}

\begin{figure*}[t]
	\centering
	\includegraphics[width=0.97\linewidth]{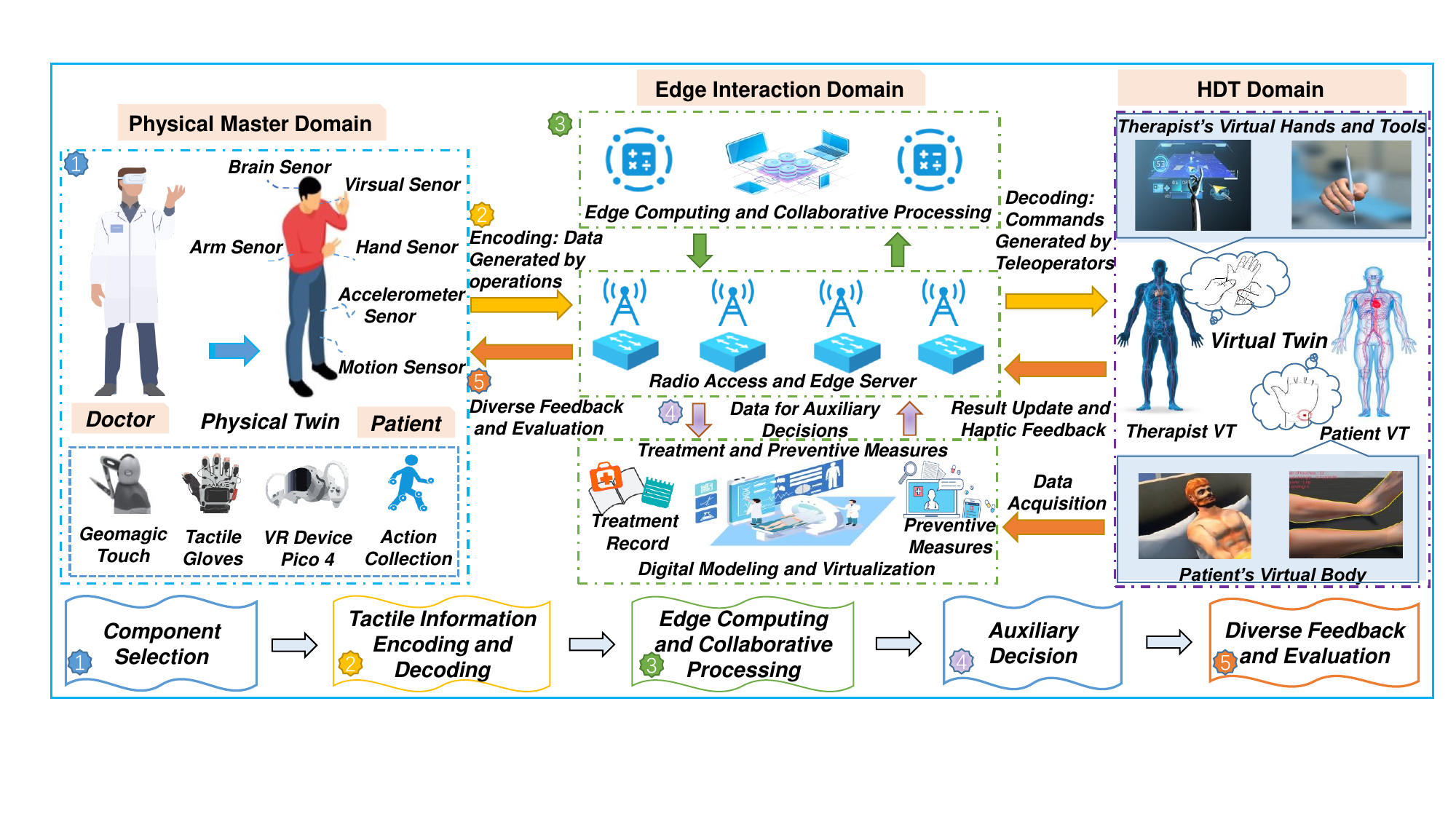}
	\caption{Key steps and core guidelines for IC-HDT-ECoTI implementations.}
	\label{Frameworks}
\end{figure*}

To potentially tackle the requirements and challenges stated in Section \ref{Framework}, we provide the key steps and core guidelines for implementing IC-HDT-ECoTI, as illustrated in Fig. \ref{Frameworks}, consisting of component selection, tactile information encoding and decoding, edge computing and collaborative processing, auxiliary decision, and diverse feedback and evaluation.
	
\subsection{Component Selection}

The successful implementation of IC-HDT-ECoTI requires a careful selection of hardware and software components. In this regard, the first step is to select appropriate components to align with the specific system requirements. Special attentions have to be paid on the interface and operating environment to ensure that the system can operate seamlessly and without introducing excessive  service latency. To obtain accurate tactile data, advanced haptic devices, such as Geomagic Touch \cite{ref1} and tactile gloves may be utilized. To ensure low latency and high reliability in communication and computation (e.g., $<$ 50 ms delay and $>$ 99.999\% reliability in haptic interaction-based rehabilitation \cite{ref9}), tactile data and VT's status information can be distributed to nearest edge servers to speed up processing. In order to provide an immersive QoE, the scene rendering can be accomplished using the unreal engine in Unity/UE, while VR glasses (e.g., Pico4) can be adopted to attain an immersive feeling. Finally, to guarantee the real-time transmission of tactile tasks and feedbacks, the time-sensitive networking (TSN) protocol \cite{ref13} may be employed as an alternative to the traditional Ethernet for further enhancing the performance.

\subsection{Tactile Information Encoding and Decoding}
\par To offer strong interactive and immersive user experiences, multimodal information, such as audio, video and tactile signals, should be transmitted between PTs and VTs in IC-HDT-ECoTI. While audio and video information are relatively straightforward to collect, tactile information usually involves complicated multimodal data, including force area, angle and size. This complexity makes the collection, encoding and decoding significantly challenging. To this end, deadband coding technology \cite{ref3}, which encodes a tactile sample only if humans can notice any changes in tactile perception, may be employed to encode the collected information into discrete values in bits for efficient transmission and data processing. The encoded information is then transmitted through the edge interaction domain to the HDT domain, where it undergoes recognition, classification and analysis to be eventually decoded. Once executed in the HDT domain, the resulting feedback is collected and encoded, and in turn transmitted back to the physical master domain for decoding. All these complete the round-trip and closed-loop interactions.

\subsection{Edge Computing and Collaborative Processing}   

The edge interaction domain is the bridge between the physical master domain and HDT domain, while the most critical node is the edge server. A large amount of tactile information is difficult to be processed and managed locally. By enabling edge computing, users can offload the data to edge servers, and thereby reducing communication bandwidth and service delay. In real-time analysis of multimodal data, distributed and collaborative approaches (e.g., distributed computing) ensures that massive amounts of information (such as tactile and video) can be processed in parallel and effectively \cite{ref14}. 
Devices, such as tactile gloves and Geomagic Touch, can also communicate and collaborate with edge servers for alleviating their computation burdens. The scheduler can apply the distributed coordination to perform the task collaboration based on the network status, node requirements and other network conditions, and further adopt end-edge collaborative processing methods to achieve data sharing between local and edge servers and among adjacent edge servers. 

\subsection{Auxiliary Decision}
\par With strong interactions between each PT-VT pair, IC-HDT-ECoTI is required to process various data on network edges in real time. For example, the tactile information of an individual, such as human skin texture and hardness, is obtained through real-time data collection. After that, the data should be analyzed to build a VT of the human body, achieving a vivid representation, including its physiology, appearance and behavioral characteristics. In addition, to assist the judgment and treatment,  IC-HDT-ECoTI can enable low-latency decision-making with real-time processing and multimodal feedback. Particularly, in massaging and hitting procedures of physical therapy, the therapist first builds a VT of the patient based on a large amount of collected historical data and update the VT by frequent information exchanging, then the therapist can use the tactile feedback during the operation (such as the amount of force and the texture of scalpel) to predict and correct the next action \cite{ref15}. Such an auxiliary decision process can also be facilitated by edge computing and collaborative processing for gaining more powerful decision-making aids, further improving the accuracy and reliability of treatments.

\subsection{Diverse Feedback and Evaluation}
\par The versatility of feedback, including video, audio and tactile signals, can offer extremely immersive QoE. However, depending on the context, different types of feedbacks may be produced and may cause different network overheads. Additionally, the development of effective performance metrics is also imperative. Unlike other systems, performance metrics of IC-HDT-ECoTI should integrate both objective and subjective evaluations. On one hand, objective metrics, such as latency, jitter, throughput, reliability and motion-to-photon delay, must be precisely measured. On the other hand, subjective metrics, focusing on capturing the realism of haptic sensations, vividness of VTs and motion sickness, have to be properly evaluated.

\begin{figure*}[t]
	\centering
	\includegraphics[width=0.87\linewidth]{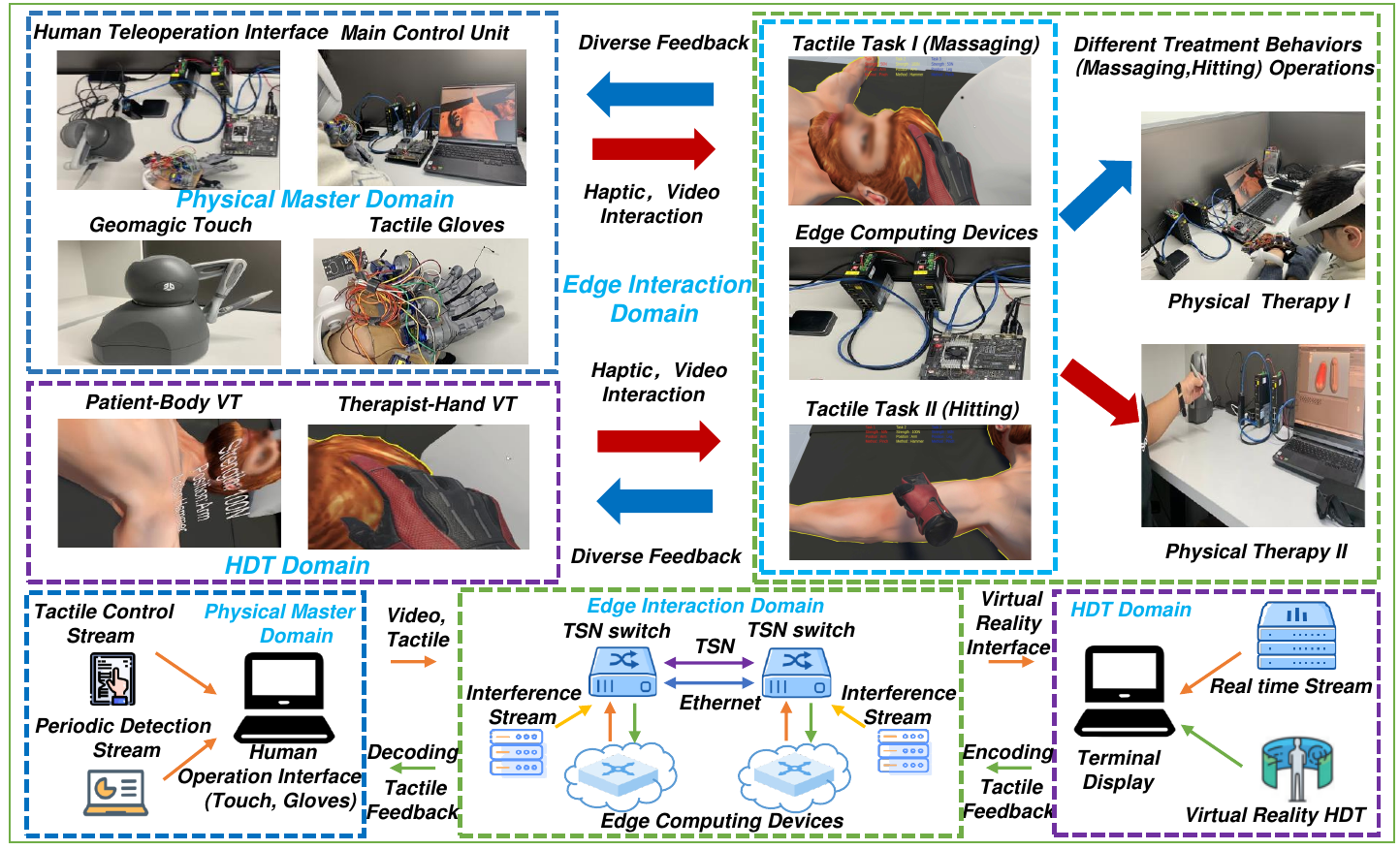}
	\caption{A case study of IC-HDT-ECoTI in physical therapy.}
	\label{case}
\end{figure*}

\section{A Case Study of IC-HDT-ECoTI}\label{Case}
In this section, we conduct a case study to demonstrate the effectiveness of the proposed IC-HDT-ECoTI in a specific application, i.e., physical therapy, following the implementation guidelines in Section \ref{Kep}. This can also be seen as a preliminary attempt to address the design requirements and challenges mentioned in Section \ref{KD}.

\subsection{Experimental Setup and System Design}

As shown in Fig. \ref{case}, we build a testbed platform by applying IC-HDT-ECoTI in physical therapy, where the physical master domain is controlled by a therapist, the HDT domain contains a hand of the therapist and a patient body, and the edge interaction domain provides both communication and computing services. The platform considers manual therapy as the primary input. A virtual hand and a virtual
body (i.e., VTs) are digitally constructed for mapping the therapist and patient (i.e., PTs) based on the perceived user data. IC-HDT-ECoTI enables the therapist to provide tactile therapy actions to the patient, allowing him/her to conduct high-density and highly interactive virtual-real operations (such as hitting and massaging) according to different situations. The platform also feeds back treatment operations to the patient, letting them to experience immersive and vivid therapies.

In the physical master domain, a Geomagic Touch and a pair of tactile gloves with VR glasses (i.e. Pico4) are adopted to initiate the tactile operations and gain the feedback, respectively. In the edge interaction domain, the embedded devices, including NVIDIA TX2 and Raspberry Pi, are utilized as edge servers for processing multimodal data in real-time, while the industrial TSN switch IE4320-10s is used to construct the network. The HDT domain consists of virtual entities in the virtual world with
action actuator, such as the physical therapist's virtual hand and the virtual patient's body that require physical therapy for realizing the digital simulation. When the system is running, the tactile information flow will be first transmitted from the physical master domain to  switches, and then  be relayed to the HDT domain. The feedbacks (e.g., tactile, video and audio) generated from the HDT domain will be transmitted back to the physical master domain via switches in return. Moreover, an interference stream generated randomly by Ostinato is introduced to simulate the background traffics in the core network for aligning with real-world scenarios.


\subsection{Performance Evaluation}

\begin{figure}[t]%
    \centering
    \subfloat[\tiny]{
        \includegraphics[width=0.47\linewidth]{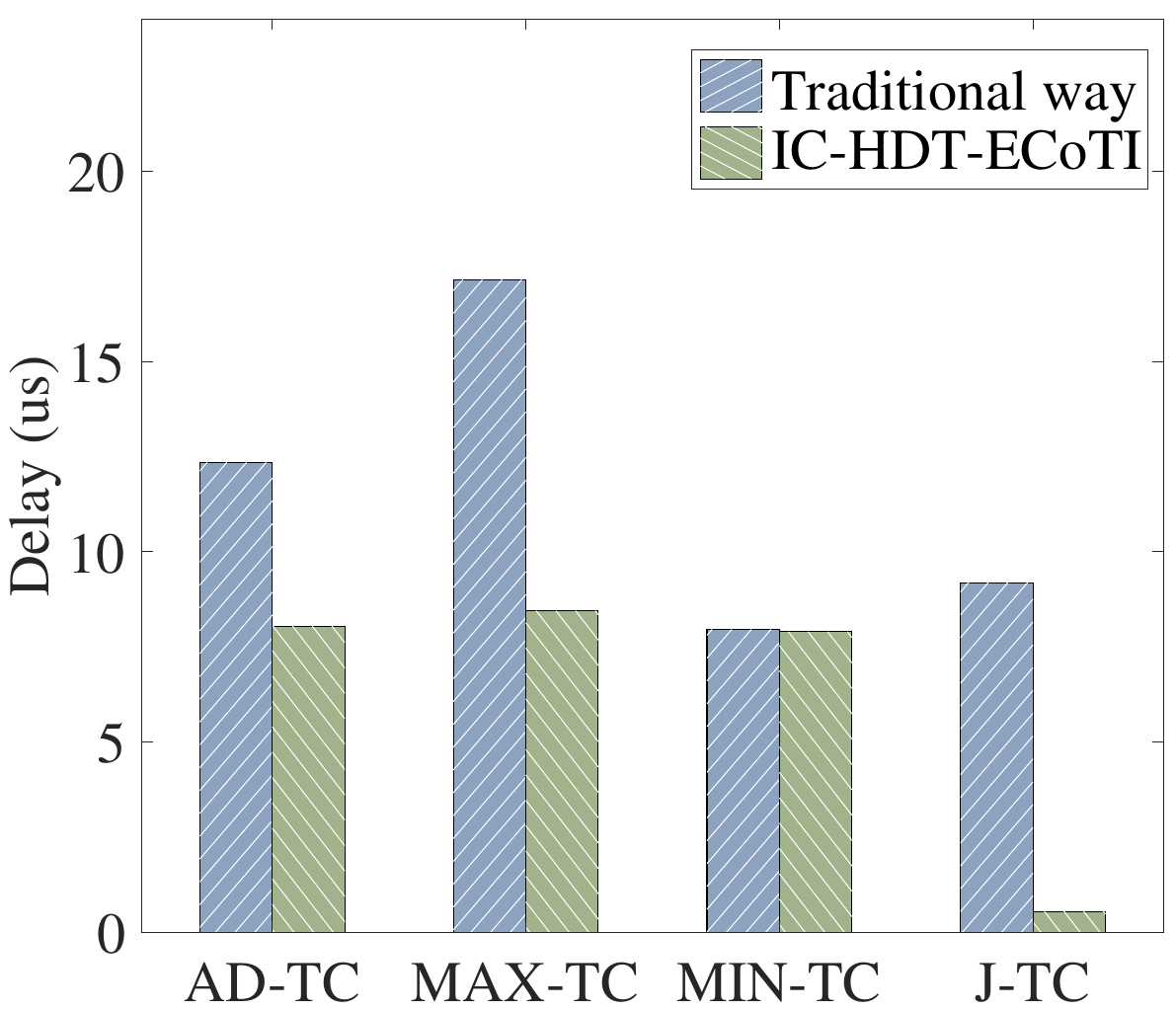} \label{subfig1}
        }\hfill \hspace{-1em}
    \subfloat[\tiny]{
        \includegraphics[width=0.47\linewidth]{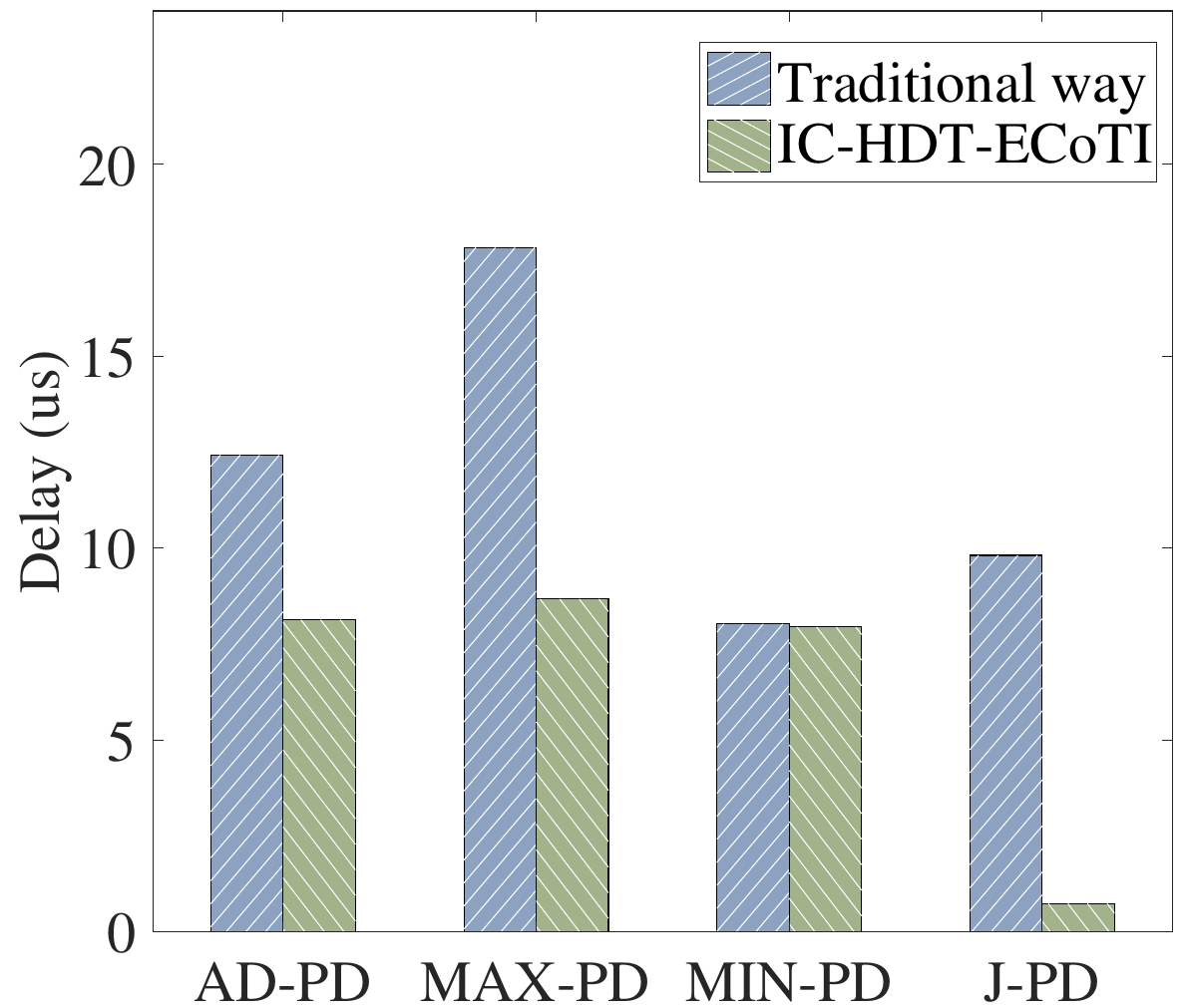} \label{subfig2}
        }
    \caption{An objective evaluation under two different frameworks.}
\end{figure}

In order to test the functions of our constructed platform, we evaluate the performance under two frameworks, i.e, i) HDT over the traditional Ethernet (namely the traditional way), and ii) the proposed IC-HDT-ECoTI. We compare their performances using three indicators: a) immersive interaction delay (TC), b) periodic stream delay (PD), and c) subjective evaluation. As shown in Fig. (\ref{subfig1}) with respect to the immersive interaction,  IC-HDT-ECoTI outperforms the traditional way in terms of the average latency (AD-TC), maximum latency (MAX-TC), minimum latency (MIN-TC) and jitter (J-TC). Fig. (\ref{subfig2}) shows the results with respect to the periodic stream, and similarly, IC-HDT-ECoTI has better performances in terms of the average delay (AD-PD), maximum delay (MAX-PD), minimum delay (MIN-PD) and jitter (J-PD).
Furthermore, we recruit 6 volunteers to test the platform subjectively. As shown in Table I, the traditional way suffers from playback delays, video interruptions and asynchronous motions, while the proposed IC-HDT-ECoTI can perform smoothly. All these indicate that, with the aim of providing strong interactions and extremely immersive QoE, our proposed framework is feasible and superior under either objective or subjective evaluations.

\begin{table}[t]
	\centering
	\caption{A subjective evaluation under two different frameworks.}
	\footnotesize
	\begin{tabular}{c|ccc}
		\hline
		\toprule
		\textbf{Framework} & \textbf{Video stream} & \textbf{Tactile feedback} & \textbf{Sync.}  \\
		\hline\toprule\toprule
		Traditional way & Choppy & Inaccurate & Jitter \\
		\hline\toprule
		 IC-HDT-ECoTI & Fluent & Responsive & Efficient \\
		\bottomrule
		\hline
	\end{tabular}
\end{table}

\par In summary, IC-HDT-ECoTI leverages its immersive communication capability and edge computing power to collect and process tactile signals in HDT system for achieving real-time communication, computation and control. For the communication resource management, the optimal routing and switching strategies are employed to ensure efficient transmission of multimodal information, enabling users to experience more realistic feedback. For the high-fidelity virtual modelling, the 3D Unity and Unreal engines are respectively utilized to model and render therapy scenes, aiming to achieve vivid services and rapid scene switching. By incorporating such visual modeling and video rendering technologies, high-definition video, accurate tactile feedback and a dynamically realistic VT based on PT can be provided. In the multimodal data analysis, IC-HDT-ECoTI adopts the fixed 1 kHz frequency deadband encoding in tactile devices, i.e., gloves and Geomagic Touch, to efficiently encode and reduce collected tactile signals to transmit in synchronization with the video data. Moreover, by placing computational and storage resources closer to end-users, collaborative edge computing guarantees low-latency and parallelism of multimodal data processing. To evaluate the overall effectiveness of physical therapy, subjective evaluation criteria from volunteers, along with objective measurements such as delay and jitter, are considered. The experimental results demonstrate that from the technical perspective, by employing the proposed IC-HDT-ECoTI, the constructed platform excels in offering strong interaction and extremely immersive QoE for HDT, while providing advantages in real-world problem-solving and practical implementations.

\section{Future Research Directions}\label{FR}

In this section, we discuss several future research directions of IC-HDT-ECoTI to potentially inspire more out-of-the-box works on this topic.

\subsection{Predictive Haptic Interaction and Resource Optimization}

Caching feedbacks on edge servers based on predictions during immersive interactions has a great potential to significantly reduce the HDT service latency. However, unlike traditional content caching, caching multi-modal signals (particularly the tactile ones) is much more challenging because most of them are hard to be modeled, and hence impossible to be properly cached. Therefore, to ensure ultra-timely and accurate responses, advanced algorithms, such as deep neural networks, along with the resource optimization are imperative.

\subsection{Security and Privacy with Human-in-the-Loop}

Security and privacy are always critical, especially for IC-HDT-ECoTI with human-in-the-loop. This may involve identifying potential vulnerabilities and threats in both physical controlled and HDT domains associated with data collection and processing, and the development of novel security and privacy mechanisms, such as differential privacy methods, to mitigate these risks. Moreover, legal, ethical and moral considerations corresponding to the implementation of HDT have to be also taken into account.

\subsection{Ultra-High Quality Modeling and User Experience} 

The profoundly immersive QoE is essential to IC-HDT-ECoTI. Enhancing QoE may involve investigating the adoption of various AI algorithms, such as tensor holography, to personalize HDT models and applications for individual users, as well as exploring the potential for integrating multi-sensory feedbacks (beyond video, audio and tactile) into HDT interfaces to create immersive and engaging user experience. This may also prompt the development of Metaverse.

\subsection{Edge Intelligence Enhanced Vivid Feedbacks}

High-fidelity engagements and interactions require vivid feedbacks. This can only be achieved if there is almost no lag (e.g., 1ms round-trip) between PTs and VTs in HDT systems. By applying edge intelligence, it is expected that complicated haptic interactions can be predicted, filtered, compressed and analyzed, such as using generative AI, fundamentally enhancing the performance of transmitting and processing feedback signals, and eventually guide IC-HDT-ECoTI to conduct active learning and self learning.

\section{Conclusion}
In this article, we study the design of an immersive communication framework for HDT by edge computing empowered tactile Internet (IC-HDT-ECoTI). We highlight major requirements and challenges, particularly in the view of demanding strong interactions and extremely immersive QoE. We present core guidelines and detailed steps for implementing such a system and conduct a case study demonstrating IC-HDT-ECoTI in physical therapy. Finally, open issues are outlined and discussed. Overall, this work may contribute to ongoing efforts towards realizing the full potential of HDT applications by applying IC-HDT-ECoTI, and pave the way for future research studies in this exciting area.

\bibliographystyle{IEEEtran}
\bibliography{IEEEabrv,TI_ref}

\vspace{-.5em}
\section*{Biography}

\small \textbf{Hao Xiang} is currently pursuing the Ph.D. degree with the College of Computer Science and Technology, Nanjing University of Aeronautics and Astronautics, China. His research interests include tactile internet, human digital twin, and edge intelligence.

\small \textbf{Changyan Yi (Member, IEEE)} is a Professor with the College of Computer Science and Technology, Nanjing University of Aeronautics and Astronautics, China. His research interests include edge computing, edge intelligence and digital twin.

\small \textbf{Kun Wu} is  currently pursuing the M.S. degree with the College of Computer Science and Technology, Nanjing University of Aeronautics and Astronautics, China. His research interests include tactile internet and edge computing.

\small \textbf{Jiayuan Chen} is currently pursuing the Ph.D. degree with the College of Computer Science and Technology, Nanjing University of Aeronautics and Astronautics, China. His research interests include edge computing and digital twin.

\small \textbf{Jun Cai (Senior Member, IEEE)} is a Professor and PERFORM Centre Research Chair with the Department of Electrical and Computer Engineering, Concordia University, Canada. His research interests include edge/fog computing and eHealth.

\small \textbf{Dusit Niyato (Fellow, IEEE)} is a President's Chair Professor with the School of Computer Science and Engineering, Nanyang Technological University, Singapore. His research interests include edge intelligence, machine learning and incentive mechanism design.

\small \textbf{Xuemin (Sherman) Shen (Fellow, IEEE)} is a University Professor with the Department of Electrical and Computer Engineering, University of Waterloo, Canada. His research focuses on network resource management, wireless network security, social networks and vehicular ad hoc networks. He is a Canadian Academy of Engineering Fellow, a Royal Society of Canada Fellow, and a
Chinese Academy of Engineering Foreign Fellow.
\end{document}